\documentclass{elsart3}

\usepackage{amssymb}

\usepackage{graphicx}

\begin{document}

\begin{frontmatter}

\title{Voltage rectification   in two dimensional Josephson junction arrays}

\author{Ver\'onica I. Marconi}

\address{Institut de physique, Universit\'e de Neuch\^{a}tel,
 CH-2000 Neuch\^{a}tel, Switzerland.}

\begin{abstract}

 We study numerically the directed motion of vortices (antivortices) under an applied ac bias in 
two-dimensional Josephson  junction arrays (JJA) with an asymmetrically modulated periodic vortex pinning
potential. We find that the  ratchet effect in large 2D JJA can be  obtained using the  RSJ model for the
overdamped vortex dynamics.   The rectification effect shows a  strong dependence on  vortex density  as 
well as an inversion of the vortex flow direction with the ac amplitude, for a wide range of high magnetic
field around $f=1/2$ ( $f$ being the vortex density).   Our results are in good agreement with very recent 
experiments by  D.E. Shal\'om and H. Pastoriza [Phys. Rev. Lett. {\bf 94}, 177001, (2005)]. 
\end{abstract}

\begin{keyword}
 Josephson junction arrays \sep ratchet effect  \sep periodic pinning 
\PACS 74.25.Qt
\end{keyword}
\end{frontmatter}

\section{INTRODUCTION}
 The study of ratchet effects in superconductors has attracted great 
interest in recent years due to the possibility of controlling vortex motion
\cite{barabasi99}. However, only very recently nano-micro engineering has allowed 
to successfully fabricate different types of asymmetric pinning potentials for the motion 
of flux quanta. Voltage rectification was spectacularly 
observed in Nb films  with a triangular dots array
of pinning potential \cite{vicent} and in Al films with square  arrays of hole
pinning sites \cite{mosch1}.

Ratchet effect has also been analyzed theoretically and  experimentally in Josephson
Junctions arrays (JJA), both in  the  classical and quantum regime, particularly for the
cases of 1D parallel JJA, SQUID's or circular arrays \cite{re_jja}. Less attention
has been paid however to larger two dimensional JJA, where collective effects are expected to
play an important role. Indeed, this collective behavior was very recently observed in 
experiments with large square JJA \cite{pastoriza}. In this case,  improved e-lithography
techniques were used to modulate the gap  between the superconducting islands and thus  an
asymmetric and periodic sawtooth potential for the vortex-antivortex motion was generated.
Voltage rectification was then clearly observed under an applied ac current  (rocking
ratchet), and  analyzed as a function of the vortex density. In  this paper we show that the
main experimental results reported in \cite{pastoriza} can be successfully  reproduced and
analyzed in detail by numerical simulations of the overdamped RSJ model for a large 2D JJA
with asymmetrically  modulated critical currents.
\section{MODEL}
We study the dynamics of asymmetrically  modulated 2D JJA using the resistively shunted junction
(RSJ) model for the SNS (superconductor-normal-superconductor) junctions in  square two
dimensional  arrays \cite{dyna}. The current flowing in the junction between two superconducting
islands  is the sum of the Josephson supercurrent and the normal current  (see details of the
model in \cite{prb2}): 
\begin{equation}
I_{\mu }({\bf n})=I_{0}\sin \theta _{\mu }({\bf n})+\frac{\Phi _{0}}{2\pi
cR_{N}}\frac{\partial \theta _{\mu }({\bf n})}{\partial t}+\eta _{\mu }({\bf %
n},t)
\end{equation}
where $I_{0}$ is the critical current of the junction between the sites $%
{\bf n}$ and ${\bf n}+{\bf \mu }$ in a square lattice [${\bf n}=(n_{x},n_{y})
$, ${\bf \mu }={\bf \hat{x}},{\bf \hat{y}}$], $R_{N}$ is the normal state
resistance and $
\theta _{\mu }({\bf n})=\theta ({\bf n}+{\bf \mu })-\theta ({\bf n})-A_{\mu
}({\bf n})=\Delta _{\mu }\theta ({\bf n})-A_{\mu }({\bf n})$
is the gauge invariant phase difference with 
$
A_{\mu }({\bf n})=\frac{2\pi }{\Phi _{0}}\int_{{\bf n}a}^{({\bf n}+{\bf \mu }
)a}{\bf A}\cdot d{\bf l}$ being  $\Phi _{0}$ the  flux quantum. The Langevin noise term $\eta _{\mu }$ 
models the contact with a thermal bath at temperature $T$ and satisfies
$ \langle \eta _{\mu }({\bf n},t)\eta _{\mu ^{\prime }}({\bf n^{\prime }},t^{\prime })\rangle
=\frac{2kT}{R_{N}}\delta _{\mu ,\mu ^{\prime }}\delta _{ {\bf n},{\bf n^{\prime }}}\delta
(t-t^{\prime })$.

The asymmetry in the underlying periodic  pinning potential is introduced by  modulating
the critical currents $I_{0}$, and therefore the Josephson coupling energy  between islands
$E_J=\Phi_{0} I _{0} / 2\pi$.
 Experimentally, the ratchet potential is generated by
modulating    the gap between the  superconducting islands, $d$, as a  sawtooth with a given
period, $p$ \cite{pastoriza}. For SNS junctions, the coupling energy decrease exponentially  with
   $d$:
  $ E_J \approx \exp[-d/\xi_N] $,  $\xi_N$ being  the coherence length in the normal metal.
 In our model we make a
simplification in order to modulate the energy couplings through the critical currents.
 We consider  a sawtooth modulation of $I_0$ where $I_0=f(n_x)$ which increases linearly from  
$I_{0_{min}}$  to $I_{0_{max}}$ with each period $p$. 
 
 An external magnetic field $H$ perpendicular to the
sample is applied, such that  
$
\Delta_{\mu}\times A_{\mu}({\bf n})=2\pi f $, where  $f=H a^2/\Phi_0$ is the vortex density and
$a$ is the array lattice spacing.  In addition, we apply  an  external ac driving, $I=I_{ac}\sin(2\pi\omega t)$ 
in the ${\bf y}$ direction.   
We take periodic boundary conditions  in both
directions.

 The dynamical equations for
  the superconducting phases are obtained after considering  conservation of the current in each node
 (see details
in \cite{prb2}). The resulting set of  Langevin  equations are solved using  a second-order
Runge-Kutta algorithm with time step $\Delta t=0.1\tau_J$ ($\tau_J=2\pi cR_N I_0/\Phi_0$) and
time integration ranges    $2.10^6 \tau_J-1.10^7\tau_J$  after the transient.   Under
ac external currents applied  perpendicularly to the sawtooth modulation,  the current-voltages
characteristics  are calculated as well as the vortex structure.  The normalization used is the
following: currents by $I_{0}$, voltages by $R_ {N}I_0$, temperature by
$I_0\Phi_0/2\pi k_B$ and frequencies by $(\tau_J)^{-1}$.  

We study the vortex rectification on ratchet-like square ($L\times L$) JJA, with period $p=8$ 
and linear modulated
critical currents between $I_{0_{min}}=0.5 I_0$ and $I_{0_{max}}=1.5 I_0$.   A wide range of  magnetic field was explored,
from the one corresponding to  only one vortex in the system ($f=1/L^2$) to very high vortex
densities. We use system sizes from $L=32$ to $L=128$.
\section{RESULTS}
\begin{figure} 
\centerline{\includegraphics[height=7.5cm]{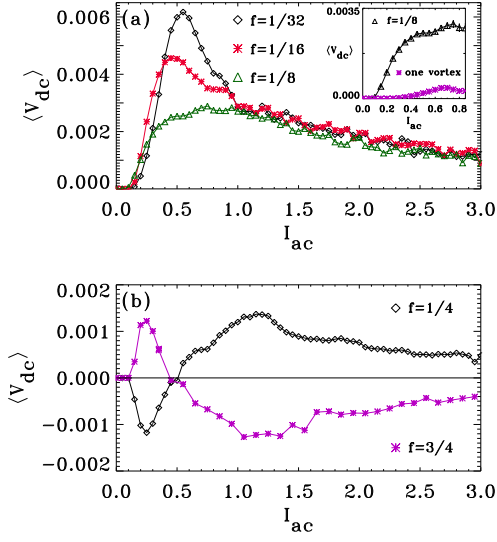}}
\caption{
Rectified vortex motion: (a) 
$I_{ac}-V$ characteristics for low vortex densities. Inset: Comparison
with one single vortex. (b) 
$I_{ac}-V$ characteristics for higher vortex densities. 
Simulations performed in $32 \times 32$ system size, at T=0.05 and $\omega=10^{-4}$. 
}
\label{fig:ivs}
\end{figure}

 It is well known that the velocity of an overdamped particle under
 an applied ac force can be rectified by introducing a periodic potential
 with broken reflection symmetry. We can thus ask 
 whether this effect, known as rocking ratchet, 
can be still observed for the motion of interacting vortices in a modulated 2D JJA.
 Since vortex velocity is proportional to voltage, 
 this kind of ratchet effect would be observable as a voltage rectification.  
In  Fig.\ref{fig:ivs}  we show clear evidence of the latter effect, i.e  
a net directional vortex motion under  the applied ac-drive of zero mean.
 We see that mean dc voltage, $\langle V_{dc} \rangle$, as a function of $I_{ac}$
 presents a maximum for an optimal value of $I_{ac}$ and decreases slowly 
with increasing $I_{ac}$. This is the  qualitative behavior  expected 
for the one particle rocking ratchet. 
    To study the role of the vortex  interactions in Fig.\ref{fig:ivs}(a)
  examples
 for different vortex densities  are presented,  $f=1/8,1/16, 1/32$.   A  decrease in  the maximum
rectification is seen while the vortex density increases as well as a shift to higher  optimal  ac
amplitude. An important feature is worthy of  being mentioned here (see  inset in Fig.1(a)): the voltage
rectification is highly negligible when only one single vortex is present in the system,
  corresponding  to
a vortex density  $f=1/(32\times32)$ in this case.  A more intriguing feature appears when the
vortex density is further increased. 
Examples are shown in Fig.\ref{fig:ivs}(b). For instance, an
  inversion of the
vortex flow  direction is observed.  These observations provide   evidence for
the relevant role of  vortex interactions. Also notice the   expected reflection symmetry around $f=1/2$ due to
the symmetries with the magnetic field of the Hamiltonian representing JJA  \cite{tink}. From  $f=0$ to
$f=1/2$ the dissipation arises from the motion of ``positive'' vortices. The opposite occurs for $f$ between
$f=1/2$ to $f=1$ where a flux of negative  vortices (antivortices)  gives a voltage response of opposite sign.
In addition to reproducing the experiments of Ref. \cite{pastoriza},  
we have observed that the instantaneous vortex lattice dynamics behind this phenomenon is very complex \cite{fut} compared with the vortex motion observed in Ref.\cite{vicent}.
          
\begin{figure} 
\centerline{\includegraphics[height=7.5cm]{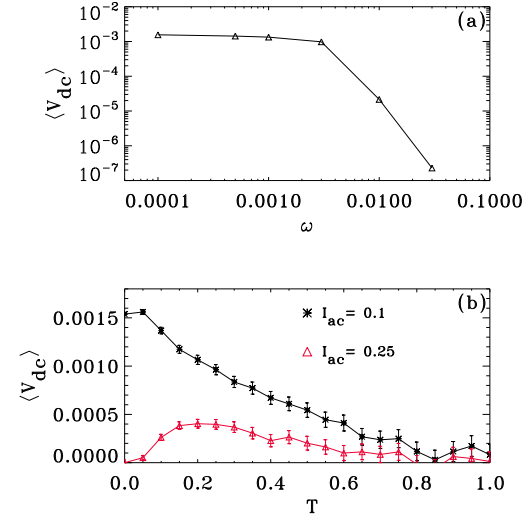}}
\caption{
Rectified dc voltage response dependences: (a) On frequency, $\omega$, 
for $f=1/8$ and $I_{ac}=0.25$. (b) On temperature, $T$,   for $I_{ac}=0.1$ and $I_{ac}=0.25$, 
both examples at $\omega=10^{-4}$.
}
\label{fig:vswT}
\end{figure}
\begin{figure}
\centerline{\includegraphics*[height=8.0cm]{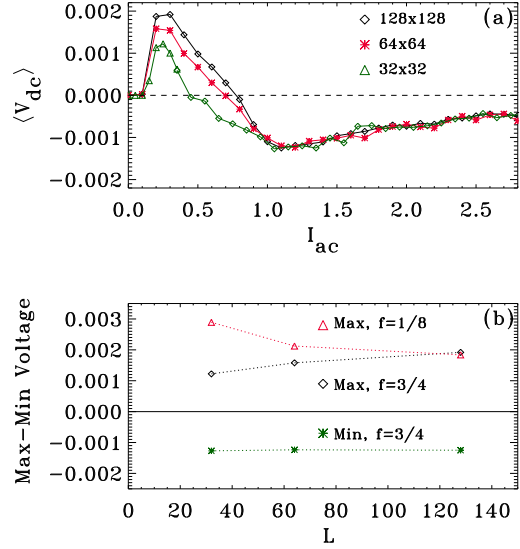}}
\caption{Size analysis:  
(a) Current-voltage characteristics for a high vortex density  ($f=3/4$) and
 different system sizes  $L=32,64,128$. 
(b) Maximum and minimum optimum mean voltages  vs system size for $f=3/4$ and $f=1/8$.}
\label{fig:se}
\end{figure}
\begin{figure}
\centerline{\includegraphics[height=7.5cm]{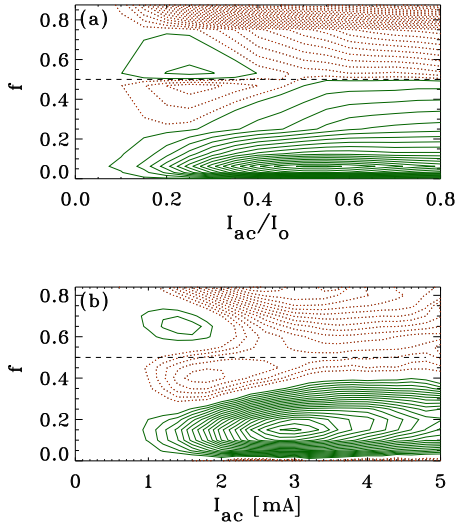}}
\caption{Rectified voltage vs $f$ and ac drive amplitude. Red dotted lines represent $\langle V\rangle_{dc} < 0$ and green solid lines, $\langle V \rangle_{dc} > 0$.
 (a) Simulations at $T=0.05$ and $\omega=10^{-4}$ in $32 \times 32$ arrays with p=8. (b)
 Experiments from \cite{pastoriza} at T=3.8K and $2\pi\omega=1 kHz$.}
\label{fig:cont}
\end{figure}

It is important to  accurately determine the range of parameters where vortex motion   rectification is
observable. In agreement with experiment, we obtain  almost zero voltage rectification for
all  ac amplitudes if the frequencies are larger that $0.001 (\tau_J)^{-1}$ due to the fast changes in
potential slope not allowing vortices to have time to explore the asymmetric underlying
potential during a  cycle. A plateau of saturation is obtained for  very low frequencies (
see Fig.  \ref{fig:vswT}(a) where an example is shown for $f=1/8$ and $I_{ac}=0.25$).   Most of
the results shown here were also obtained at zero temperature  as expected for 
rocking ratchets. It is also useful to
explore the range of temperatures where rectification  occurs. An analysis
is shown in Fig. \ref{fig:vswT}(b) for two  values of the ac amplitude which are  representative of the
 behavior  
of almost all the vortex densities studied (see also  Fig.1): $I_{ac}=0.1$ and $I_{ac}=0.2$.   

In  Fig. \ref{fig:se}(a) we show that our main results at  high vortex density ($f=3/4$) are 
independent of 
system size.  The key features, such as  vortex rectification 
and current reversal effects do not depend strongly on system sizes. An analysis 
for the maximum and minimum values of the rectification vs L is shown in Fig. \ref{fig:se}(b)
 for two vortex densities,
 $f=1/8$ and $f=3/4$ (for more
detailed explanation on the dynamics at intermediate ac amplitudes see \cite{fut}).        

Finally, in Fig. \ref{fig:cont}, contour plots show the occurrence of voltage rectification 
in the $I_{ac}-f$ plane.  The plot was generated from $\langle V \rangle_{dc}$ vs $I_{ac}$ curves obtained 
for different magnetic fields as those shown in Fig. 1  and Fig. 3(b). A qualitatively good agreement 
with experiments \cite{pastoriza} is obtained.  A more detailed inspection shows some differences comparing 
both numerical-experimental data. The denser and softer experimental contours are obtained by continuously
changing
  the magnetic field while the simulations were performed for a discrete number of vortex densities. 

 In conclusion,
 we find  rocking ratchet effects in  
asymmetric modulated 2D JJA  using the  RSJ model for overdamped vortex dynamics.  
 A  strong dependence on  vortex density  as  well as an inversion of the vortex flow is found, 
in good  agreement with experiments.
 
We acknowledge useful discussions with  D. Dom\'{\i}nguez, P. Martinoli,
 D.E Shal\'om (also for experimental data shown), H. Pastoriza and A.B. Kolton.   
This work was supported by the Swiss National Foundation.

\end{document}